\documentclass[preprint,english,showpacs]{revtex4}

\usepackage{babel,amsmath,amssymb}
\usepackage[dvips]{graphicx}
\begin{document}
\newcommand{\be}{\begin{equation}}
\newcommand{\ee}{\end{equation}}
\newcommand{\bear}{\begin{eqnarray}}
\newcommand{\eear}{\end{eqnarray}}
\newcommand{\dpf}{\frac}

\title{ Probing the curvature of the Universe from supernova measurement}
\author{Bin Wang}
\email{binwang@fudan.ac.cn}
\affiliation{ Department of Physics, Fudan University, Shanghai 200433, China}
\author{Yungui Gong}
\email{gongyg@cqupt.edu.cn} \affiliation{Institute of Applied
Physics and College of Electronic Engineering, Chongqing
University of Posts and Telecommunications, Chongqing 400065,
China}
\author{Ru-Keng Su}
\email{rksu@fudan.ac.cn}
\affiliation{ China Center of Science and Technology (World Lab), Beijing 100080 and Department of Physics, Fudan University, Shanghai 200433, China}

\begin{abstract}
We study the possibility to probe the spatial geometry of the
Universe by supernova measurement.  We illustrate with an accelerating universe
model with infinite-volume extra dimensions, for which the
1$\sigma$ level supernova results indicate that the Universe is
closed.
\end{abstract}

\pacs{ 04.50.+h, 98.80.Cq}
\maketitle

The precision measurements of the Wilkinson Microwave Anisotropy
Probe (WMAP) have provided high resolution Cosmic Microwave
Background (CMB) data \cite{1,2} and elevated cosmology to a new
maturity. Among interesting conclusions that have been reached
from these data, the WMAP results indicate that while flatness of
the Universe is confirmed to a spectacular precision on all but
the largest scales \cite{1}, a closed universe with positively
curved space is marginally preferred \cite{3,4,5,6,7}. This
tendency of preferring closed universe is not restricted to the
WMAP data, it appeared in a suite of CMB experiments before
\cite{8,9,10}. The improved precision from WMAP provides further
confidence.

In addition to CMB, recently it was argued that the cubic
correction to the Hubble law measured with high-redshift
supernovae is another cosmological measurement that probes
directly the spatial curvature \cite{11}. This is the first
non-CMB probe of the spatial geometry, which can provide a
cross-check to the result got by CMB. In a toy model, it was
already found that a curvature radius is larger than the Hubble
distance \cite{11}.

Our Universe is accelerating  rather than decelerating. This may
be regarded as the evidence for a nonzero but very small
cosmological constant (see \cite{12} for a review and related
references in \cite{13}). Another possibility is that the
phenomenon of accelerated expansion is caused by a breakdown of
the standard Friedmann equation due to the extra-dimensional
contribution \cite{14,15,16,17}. Studies on this possibility can
also be found in \cite{18}. In this work we will consider the
accelerated universe model resulted from the gravitational leakage
into extra dimensions \cite{16}. We will attempt to extract
information from the full redshift data to test the spatial geometry.

Consider the accelerating universe described by the model with
infinite-volume extra dimensions \cite{16}, the Friedmann equation
is expressed as
\be    
H^2+\dpf{k}{a^2}=\left\{ \sqrt{\dpf{\rho}{3M_p ^2}+\dpf{1}{4r_c
^2}}+\dpf{1}{2r_c} \right\}^2,
\ee
where $\rho$ is the total
cosmic fluid energy density and $r_c$ is the crossover scale. Eq.
(1) can also be recasted in terms of the redshift as
\be     
H^2=H_0^2\left\{-\Omega_k(1+z)^2+\left[\sqrt{\Omega_{r_c}}+\sqrt{\Omega_{r_c}+\Omega_M
(1+z)^3}\right]^2\right\}, \ee where $\Omega_{r_c}=\dpf{1}{4r_c ^2
H_0 ^2}$, $\Omega_M$ is the non-relativistic matter density. The
conservation for energy-momentum tensor of the cosmic fluid is
still described by
\be     
\dot{\rho}+3H(\rho+P)=0.
\ee

Using definitions $q_0=-\dpf{\ddot{a} a}{\dot{a}^2}\vert_0,
j_0=\dpf{\dddot{a} a^2}{\dot{a}^3}\vert_0$ with dot denoting the
differentiation with respect to time $t$  for the
deceleration parameter and the ``jerk", respectively, we have directly from equation (2)
\bear    
q_0 & = & -(1+\Omega_{k_0})+\dpf{3\Omega_M\sqrt{1+\Omega_{k_0}}}{2\sqrt{\Omega_{r_c}+\Omega_M}}, \\
j_0 & = &
(1+\Omega_{k_0})-\dpf{9\Omega_M^2\sqrt{\Omega_{r_c}}}{4(\Omega_{r_c}+\Omega_M)^{3/2}},
\eear
where the normalization of (2) at the present epoch
\be     
\Omega_{r_c}=\dpf{(1+\Omega_{k_0}-\Omega_M)^2}{4(1+\Omega_{k_0})}
\ee
has been employed. With (6), $q_0$ and $j_0$ are only determined by $\Omega_M$ and $\Omega_{k_0}$.

The physically reasonable cosmic model has the following
requirements \cite{19}: (1) the total density is currently not
increasing as a function of time; (2) for causality and stability,
the present sound speed $c_s$ of the total system satisfies $0\leq
c_s^2\leq 1$.

Employing (1) and (3), the variation of the total cosmic fluid
energy density and the sound speed of the total cosmic fluid at
the present epoch are
\be    
\dot{\rho}\vert_0  =  -6M_p ^2
H_0^3(1+q_0+\Omega_{k_0})\left[1-\dpf{\sqrt{\Omega_{r_c}}}{\sqrt{1+\Omega_{k_0}}}\right],
\ee
\bear
c_s^2 &  = & \dpf{\dot{P}}{\dot{\rho}}\vert_0 \nonumber \\
& = &
\dpf{(j_0-1-\Omega_{k_0})(1-\sqrt{\Omega_{r_c}}/\sqrt{1+\Omega_{k_0}})+(q_0+1+\Omega_{k_0})^2\sqrt{\Omega_{r_c}}/(1+\Omega_{k_0})^{3/2}}{3(1+q_0+\Omega_{k_0})[1-\sqrt{\Omega_{r_c}}/\sqrt{1+\Omega_{k_0}}]}.
\eear

The first requirement implies
\be  
(1+q_0+\Omega_{k_0})(1-\dpf{\sqrt{\Omega_{r_c}}}{\sqrt{1+\Omega_{k_0}}})\geq
0. \ee Employing (6) and the fact that $\vert\Omega_{k0}\vert\leq
0.1$ as a consequence of CMB data, the above requirement reduces
to
\be      
1+q_0+\Omega_{k_0}\geq 0. \ee Using (4), we see that Eq. (10) can
obviously be satisfied.

The second requirement can now be written in a simplified form as
\be         
f_1\leq j_0\leq f_2, \ee where
$f_1=(1+\Omega_{k_0})-\dpf{(1+\Omega_{k_0}-\Omega_M)(q_0+1+\Omega_{k_0})^2}{(1+\Omega_{k_0})(1+\Omega_{k_0}+\Omega_M)}$
and
$f_2=4(1+\Omega_M)+3q_0-\dpf{(1+\Omega_{k_0}-\Omega_M)(q_0+1+\Omega_{k_0})^2}{(1+\Omega_{k_0})(1+\Omega_{k_0}+\Omega_M)}$.
Substituting Eqs. (4)
and (6) into the expression of $f_1$, we find that $j_0=f_1$,
which means that the sound speed of the total system in this model
is exactly zero.

We now turn to determine the cosmological density parameters from
the supernova (SN) Ia data compiled by Riess et al. \cite{20}. The
likelihood for the parameters $\Omega_M$ and $\Omega_{k_0}$ can be
obtained from a $\chi^2$ statistics \cite{20,21}, where \be
 \chi^2(H_0,\Omega_M, \Omega_{k_0})=\sum_i \dpf{[\mu_{p,i}(z_i, H_0, \Omega_{k_0},
 \Omega_M)-\mu_{o,i}]^2}{\sigma_i^2},
\ee
$\mu_p=5\log_{10}(d_L/{\rm Mpc})+25$ and $\mu_o$ are distance
modulus for the model and the observations, respectively.
$d_L$ is  the luminosity distance defined for the Friedmann-Robertson-Walker universe model as
\begin{eqnarray}
d_L & = & a_0 (1+z)r_1 \\
    & = & \left \{
    \begin{array}{ll}
    a_0 (1+z)\sin [\dpf{1}{a_0 H_0}\int^z _0\dpf{dz'}{E(z')}], & {\rm closed} \nonumber \\
    \dpf{(1+z)}{H_0}\int^z _0\dpf{dz'}{E(z')}, &{\rm flat} \nonumber \\
    a_0 (1+z)\sinh [\dpf{1}{a_0 H_0}\int^z _0\dpf{dz'}{E(z')}], & {\rm open} \nonumber
    \end{array}
    \right.
\end{eqnarray}
for closed, flat and open universes respectively.  The
 function $E(z)$ quantifies the expansion rate as a
function of redshift defined as $H(z)=H_0 E(z)$. $\sigma_i$ in
(12) is the total uncertainty in the observation. Marginalizing
our likelihood function over the nuisance parameter $H_0$ by
integrating the likelihood function $L=\exp(-\chi^2/2)$ over all
possible values of $H_0$ with a flat prior assumption on $H_0$,
yields the confidence intervals shown in Fig. 1 by combining Eqs.
(2), (6), (12) and (13).
\begin{figure}[htb]
\begin{center}
\includegraphics[width=0.7\textwidth]{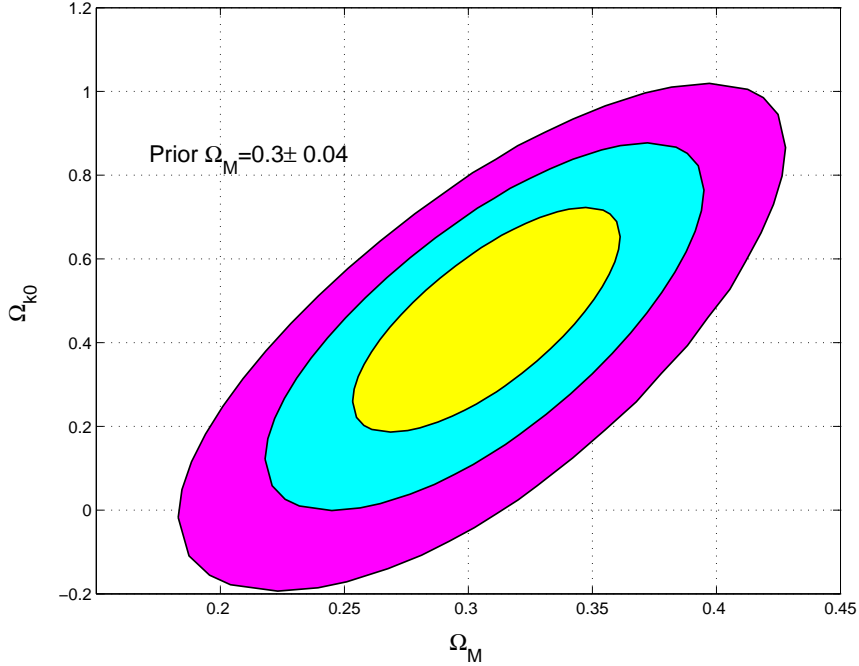}
\end{center}
\vspace*{-0.3in} \caption{The $1\sigma$, $2\sigma$ and $3\sigma$
confidence contours for $\Omega_M$ and $\Omega_{k0}$ with the
prior $\Omega_M=0.3\pm 0.04$ \cite{tegmark}.}
\end{figure}

Using the contour $\Omega_{k_0}, \Omega_M$ values, we can get the
corresponding $q_0, j_0$ and $f_1$ as plotted in Fig. 2. Note that
the contours shown here are from the gold sample SN Ia data
compiled in \cite{20}.
\begin{figure}[htb]
\begin{center}
\includegraphics[width=0.7\textwidth]{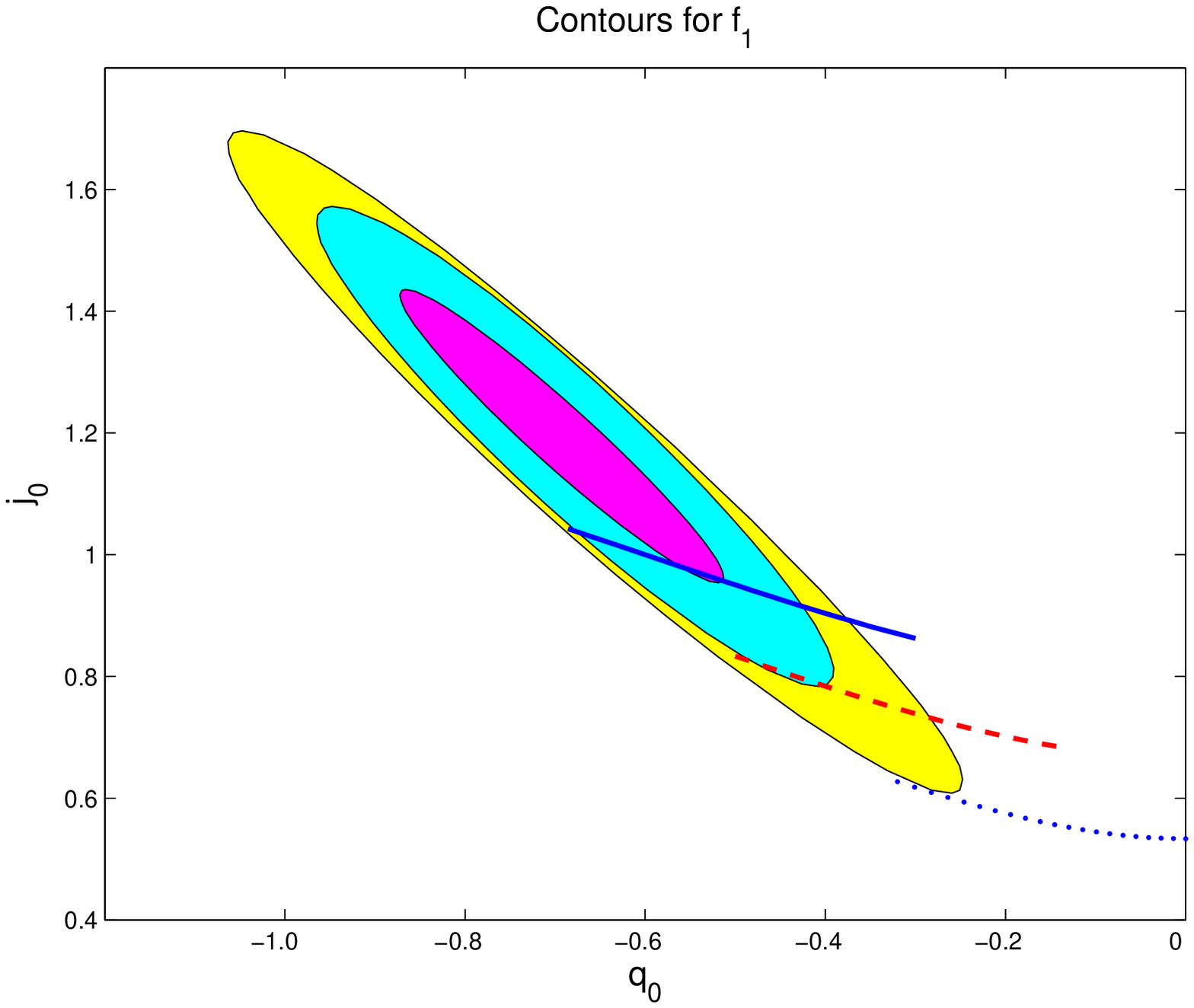}
\end{center}
\vspace*{-0.3in} \caption{The $1\sigma$, $2\sigma$ and $3\sigma$
confidence contours for $q_0$ and $j_0$. The solid, dashed and dotted lines are plots for $j_0=f_1$ with $\Omega_M$ varying in the range
[0.2-0.4] and $\Omega_{k_0}=0.2, 0, -0.2$ respectively.}
\end{figure}
\begin{figure}[htb]
\begin{center}
\includegraphics[width=0.7\textwidth]{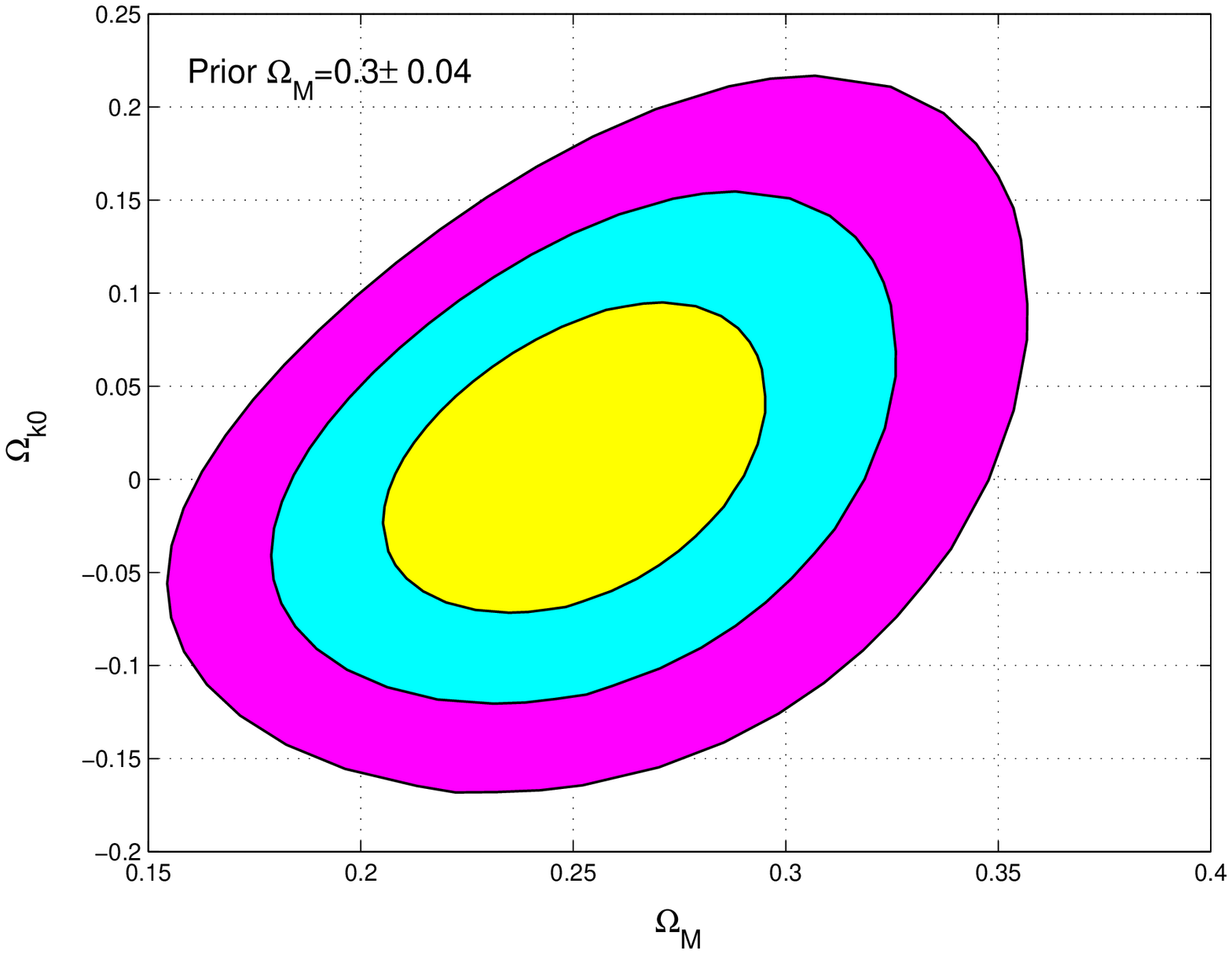}
\end{center}
\vspace*{-0.3in} \caption{The $1\sigma$, $2\sigma$ and $3\sigma$
confidence contours for $\Omega_M$ and $\Omega_{k0}$ by combining the CMB data and supernova data with the
prior $\Omega_M=0.3\pm 0.04$.}
\end{figure}

Lines added in Fig.2 show the second requirement for a reasonable
cosmic model, $j_0=f_1$, with $\Omega_M$ varying in the range
[0.2-0.4] and different $\Omega_{k_0}$ for open, flat and closed
universes, respectively. It is clear that in the 2$\sigma$ level,
there are only overlaps with the supernova data for
$\Omega_{k_0}>0$.  This corresponds to say that the data favors
the closed universe almost at 2$\sigma$ level.

To obtain tighter constraints on the parameter space, we also
include constrains from combined WMAP data \cite{1,2} and SN Ia
data. We minimize
\begin{equation}
\label{lrmin}
\chi^2=\sum_i{[\mu_{p,i}(z_i,H_0,\Omega_{k0},\Omega_M)-\mu_{o,i}(z_i)]^2\over
\sigma^2_i}+{[\mathcal{R}_p(\Omega_{k0},\Omega_M)-\mathcal{R}_o]^2\over
\sigma^2_\mathcal{R}},
\end{equation}
where $\sigma_\mathcal{R}$ is the uncertainty in $\mathcal{R}$,
the CMB shift parameter $\mathcal{R}\equiv \Omega^{1/2}_M H_0
r_1(z_{\rm ls})=1.710\pm 0.137$ \cite{shift} and $z_{\rm
ls}=1089\pm 1$ \cite{1,2}. The results are shown in Fig.3. The
combined constraints give $\Omega_M=0.25^{+0.05}_{-0.04}$ and
$\Omega_{k0}=0.01^{+0.09}_{-0.08}$. This shows that in the absence
of positive spatial curvature, $\Omega_M$ tends to take a smaller
value. It implies that from the observed $\Omega_M$ around $0.3$,
we should have the positive curvature.

From the SN Ia data, $\Omega_{r_c}$ is constrained to be $0.23$
($[0.18, 0.28]$ in $1\sigma$ region; $[0.14, 0.31]$ in $2\sigma$
region and $[0.1, 0.33]$ in $3\sigma$ region); combined with CMB,
we have tighter constraint, $\Omega_{r_c}=0.14$ ($[0.12, 0.16]$ in
$1\sigma$ region; $[0.11, 0.17]$ in $2\sigma$ region and $[0.10,
0.18]$ in $3\sigma$ region). The corresponding crossover scale
$r_c=1.04 H_0^{-1}$ from supernova data and $r_c=1.34H_0^{-1}$
from combined CMB and SN Ia data. This constrained parameter is in
good agreement with the result comes from lunar laser ranging
experiments that monitor the moon's perihelion procession with a
great accuracy \cite{Dvali}.

In summary, we have probed the geometry of a specific model
describing the accelerating  universe by using the full redshift
data in supernova measurements. To almost 2$\sigma$ level, our
result indicates that the universe is closed. This result is also
favored by including WMAP data constraint,  which agrees to a
suite of CMB experiments. The result obtained is consistent with
the interpretation from other models, e.g. the matter plus
cosmological constant case, that the Riess et al. data show a
tendency towards a closed universe. Of course it is too early to
draw conclusions just on 2$\sigma$ level data, and we expect that
future supernova measurements can determine the spatial curvature
precisely.

ACKNOWLEDGHEMENT: This work was partically supported by  NNSF, China,
Ministry of Science and
Technology of China under Grant No. NKBRSFG19990754 and Ministry of
Education of China.
Y. Gong's work was supported by Chongqing University of
Post and Telecommunication under grant Nos. A2003-54
and A2004-05.
R. K. Su would like
to acknowledge National Basic Research Program of China 2003CB716300. B. Wang thanks helpful discussions with Prof. E. Abdalla.

\end{document}